\documentclass[prd,twocolumn,nofootinbib,superscriptaddress]{revtex4-2}
\usepackage{graphicx}
\usepackage{xcolor}

\usepackage{preamble}

\begin{document}

\title
{No hair but plenty of feathers: are birds black holes?}

\author{Andrew Laeuger \orcidlink{0000-0002-8212-6496}}
\affiliation{TAPIR, California Institute of Technology, Pasadena, CA 91125, USA}
\affiliation{LIGO Laboratory, California Institute of Technology, Pasadena, CA 91125, USA}

\author{Taylor Knapp \orcidlink{0000-0001-8474-4143}} 
\affiliation{TAPIR, California Institute of Technology, Pasadena, CA 91125, USA}
\affiliation{LIGO Laboratory, California Institute of Technology, Pasadena, CA 91125, USA}

\date{April 1, 2026}

\begin{abstract}
    The imitative verb ``chirp" is thought to originate from 16th-century Middle English. 
    Meanwhile, this same word has been used to describe the gravitational waves (GWs) emitted from the merger of compact objects, such as black holes and neutron stars, since at least the 1990s. 
    Motivated purely by this linguistic overlap, we study whether the chirps of birds can be modeled by compact binary waveforms. 
    In particular, we consider a test case of the Northern cardinal (\textit{Cardinalis cardinalis}), finding that its time-reversed chirp can be approximately modeled by that of a high mass ratio, precessing black hole binary, with a number of indications towards extreme matter effects or beyond the Standard Model physics.
    Importantly, this waveform correspondence is not so straightforward for all bird species, as some chirp morphologies are far more akin to glitches seen in GW observatories.
    With these comparisons made, we propose an alternative solution to the longstanding philosophical conundrum: rather than the chicken or the egg, perhaps it was the Big Bang which truly came first. 
\end{abstract}

\maketitle

\section{Introduction}
The evolutionary history of birds has been thoroughly investigated since the discovery of an \textit{Archaeopteryx} dinosaur fossil in 1861 by Hermann von Meyer \cite{Meyer1861}. 
In the 165 years since then, the avian evolutionary biologist community has arrived at the consensus that modern day birds evolved from feathered theropod dinosaurs in the Late Jurassic period, roughly 150 to 165 million years ago \cite{Prum2015, Brusatte2015}. 
Yet, even after over a century and a half of fervent experimental research, incorporating state of the art gene sequencing technologies and producing countless peer-reviewed articles, a key question remains:

\vspace{0.25cm}
\begin{center}
\textit{What if all of it is wrong?}
\end{center}
\vspace{0.25cm}

Indeed, the true nature of these winged creatures has come into question in recent years, with many individuals parroting (perhaps satirically) the conspiracy theory that the United States government replaced all birds with drone copies in the 1960s in attempt to establish a mass citizen surveillance network. 
These claims often point to studies conducted during the COVID-19 pandemic and periods of government shutdown which report decreased bird sightings by citizen scientists during these periods as supporting evidence \cite{basile2021birds}. 

We believe that these ideas are deeply misguided; instead, we propose that the true answer may lie down a completely unexpected path: gravitational wave astronomy. 
The coalescence of a compact binary composed of black holes or neutron stars generates a gravitational waveform which gradually rises in both amplitude and frequency before quickly dying off.
This characteristic CBC waveform is often referred to as a ``chirp", with this term first appearing in the literature in the 1990s to describe relevant physical quantities, most notably the ``chirp mass" \cite{Kafka1988, Cutler1992, 2024AAS...24321604B}.
Sonifications of GW signals have played an integral role in communicating the scientific results of the GW astronomy community since the detection of GW150914 \cite{LIGOScientific:2016aoc} over a decade ago.
Yet, in the authors' experiences, presenters rarely, if at all, choose to imitate a ``chirp" when mimicking a compact merger, instead opting for sounds closer to a ``whoop" or ``brrrp".

\begin{figure}
    \centering
    \includegraphics[width=\linewidth]{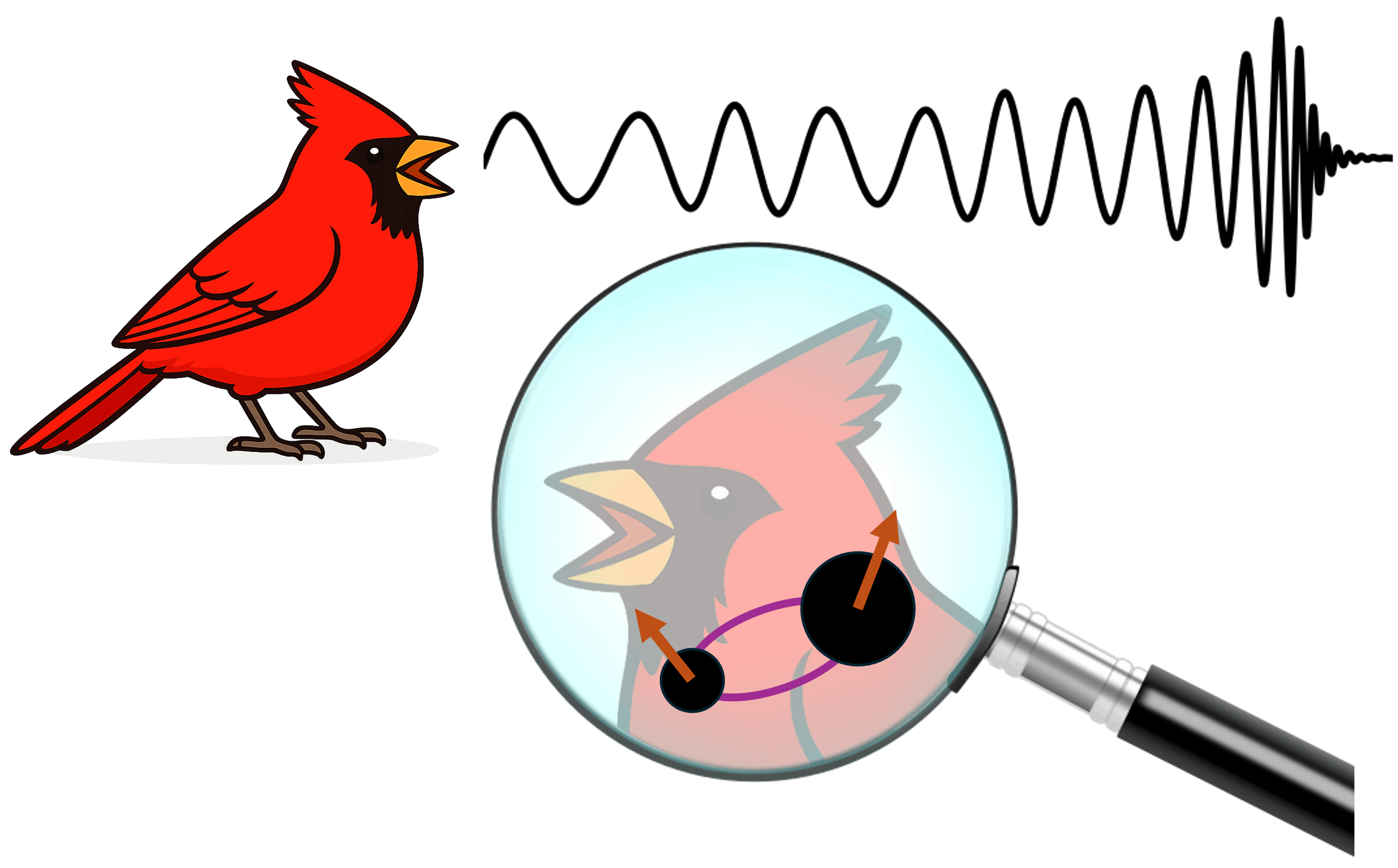}
    \caption{The central tenet of this paper: careful investigation of the calls of certain birds may reveal that these creatures actually conceal compact binaries whose radiated gravitational waves match the ``chirps" we observe in nature.}
    \label{fig:birbbh}
\end{figure}

The fact that the GW scientific community has settled on a piece of technical jargon identical to a word heavily associated with the calls and songs of birds, despite almost never invoking it in practice, appears to us far too coincidental to not warrant a thorough investigation of this discontinuity. 
In this work, we complete such an investigation, presenting a rudimentary case study wherein we outline key tools essential to revealing the dark nature of the birds in flight around us all.
We focus specifically on the Northern cardinal (\textit{Cardinalis cardinalis}), a species which is found across much of the southeastern portion of North America. 
After careful analysis on recordings of its time-reversed chirp, we find that this waveform can be approximately described by a high mass-ratio precessing BBH signal.

The remainder of this manuscript proceeds as follows. In Sec. \ref{sec: methods}, we describe our procedures for obtaining and cleaning chirp data and matching it to CBC waveforms. In Sec. \ref{sec: results}, we provide some initial results from application of these methods to the Northern cardinal chirp. We also highlight the glitchy nature of certain species as a caveat to those who wish to build upon our methods. Finally, in Sec. \ref{sec: conclusion}, we outline exciting possibilities for the continued development of this new program 
in gravitational wave science and conclude. 

Bird chirp recording data was obtained from the Xeno-canto Foundation (\url{https://xeno-canto.org}), a publicly available database of wildlife sound recordings.

\section{Methods}
\label{sec: methods}

In this section, we describe our methods of isolating and fitting our chirp signal with a preferred waveform model. 
We begin with Sec. \ref{sec:spectrogram}, which outlines our approach for extracting spectral data and a candidate waveform from the bird chirp audio file, respectively. 
We choose to isolate and fit the dominant mode of the bird call spectrogram, which is discussed in Sec. \ref{sec:filtering}.
Our procedure for fitting the isolated spectral ridge with the \texttt{SEOBNRv5PHM} \cite{Ramos-Buades:2023ehm} precessing spin model is in Sec. \ref{sec:SEOBNRv5PHM}.

\subsection{Extracting the spectrogram}
\label{sec:spectrogram}

We obtain data from the Xeno-canto Foundation in the form of \texttt{.wav} audio files, which are typically sampled at a rate of 48 kHz. To extract a single chirp from the recording, we compute the signal power averaged over 50 ms intervals, locate the time of peak signal power, and window the chirp where the averaged power falls below 0.5\% of the maximum on either side of that peak time.

To further assess \textit{Cardinalis cardinalis} as a CBC candidate, we transform into the frequency domain and produce a similar spectrogram to LVK GW detections. 
To do so, we compute a Short-Time Fourier Transform (STFT) of the signal with a large FFT window ($n_{\rm fft} = 2048$), short hop length (0.167 ms), and narrow analysis window (2.67 ms).
These choices yield a high resolution across time in the time-frequency representation of the spectrogram. 
To return to the time domain and construct a waveform from the spectrogram, we compute the inverse of the STFT. 

For both the spectrogram and the time domain signal, we implement a time-reversal of the call; doing so vastly improves the morphological similarity between the bird chirp and a GW waveform. The implications of this behavior on the underlying physics are highly nontrivial, and we will explore them further in Sec. \ref{sec: conclusion}.

\begin{figure}
    \includegraphics[width=\columnwidth]{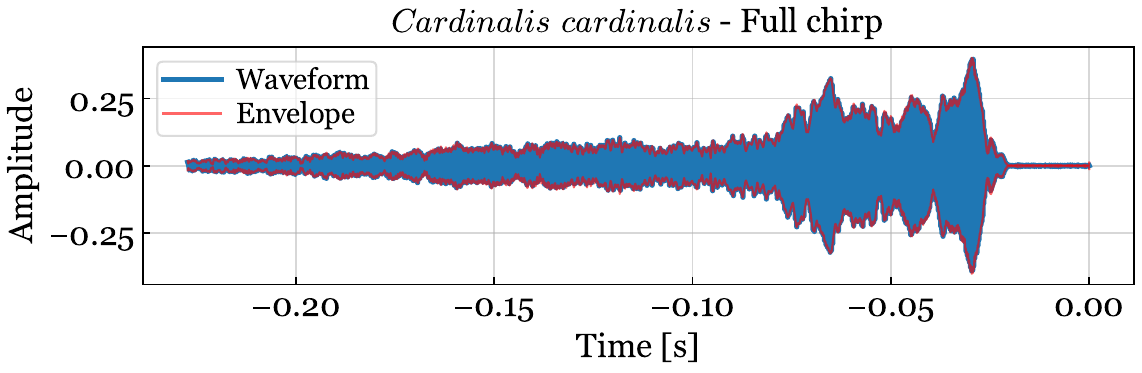}
    \includegraphics[width=\columnwidth]{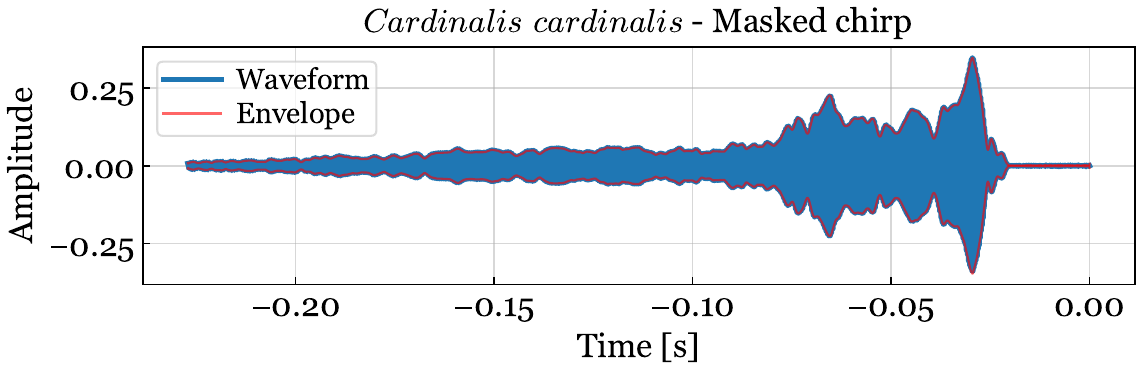}
    \caption{Waveforms obtained from the Northern cardinal time-reversed chirp. 
    The top panel gives the full unmasked bird call. 
    The bottom panel is the isolated portion of the signal corresponding to the dominant spectral track of the chirp, computed via an inverse STFT applied to our masked spectrogram in the lower panel of Fig. \ref{fig:filtered_spectrogram}.}
    \label{fig:waveforms}
\end{figure}

The time-reversed chirp of \textit{Cardinalis cardinalis} appears in the top panel of Fig. \ref{fig:waveforms}, while the result of the STFT applied to this waveform is shown in the top panel in Fig. \ref{fig:filtered_spectrogram}. The spectrogram is colored by the amplitude of the signal in each frequency bin over each time window in the STFT. 

\subsection{Primary band filtering}
\label{sec:filtering}
\begin{figure}
    \includegraphics[width=\columnwidth]{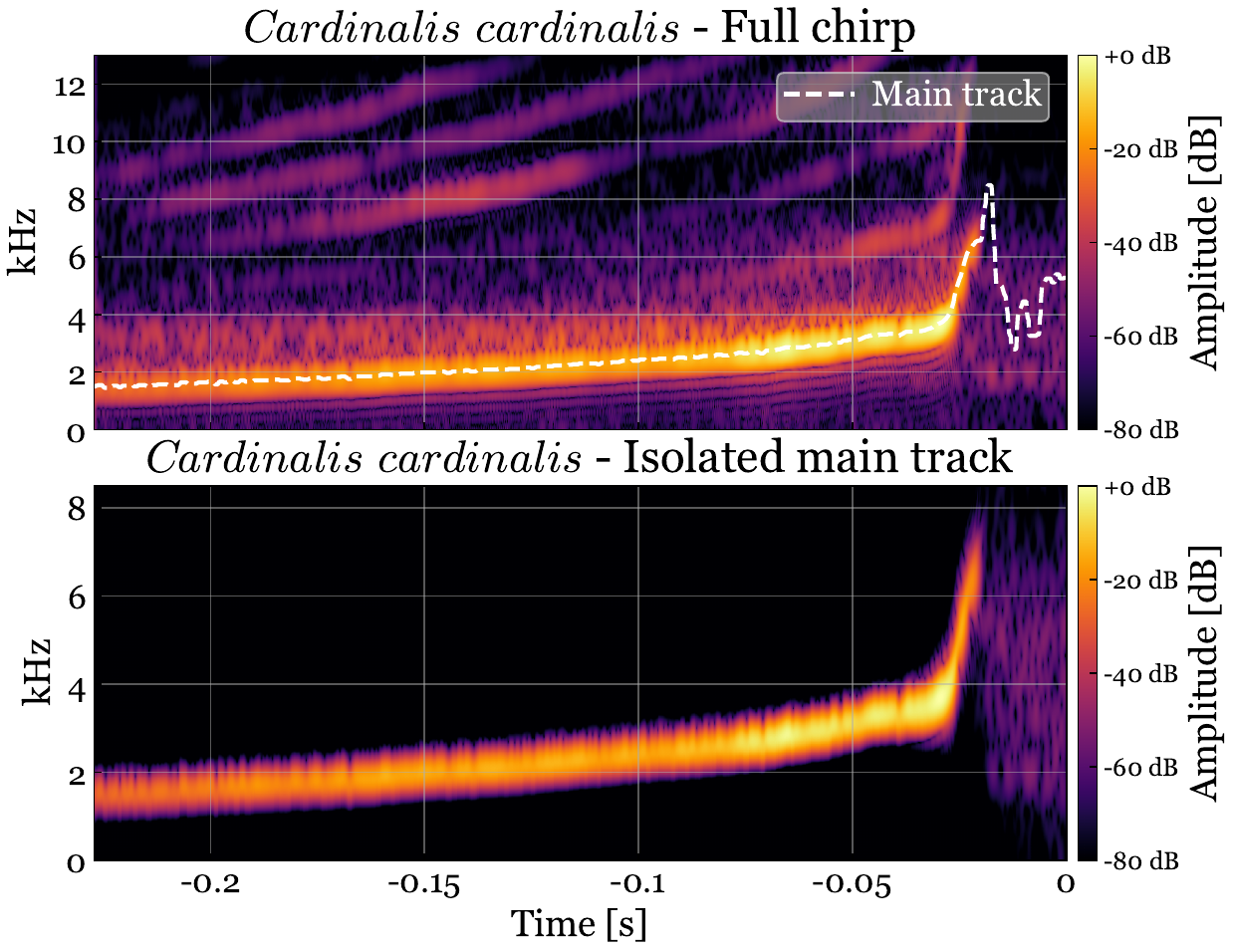}
    \caption{Spectrograms of the Northern cardinal chirp, computed via an STFT on the time-reversed waveform, where the color indicates the magnitude in each frequency bin at a given time. 
    The top panel shows the full spectrogram with the dominant spectral ridge (``main track", dashed white line) determined via our greedy algorithm.
    The bottom panel shows only the dominant spectral ridge, where we have applied a Gaussian mask which captures the full width of the track. 
        \label{fig:filtered_spectrogram}
        }
\end{figure}

Examining the unfiltered spectrogram in Fig. \ref{fig:filtered_spectrogram}, we can identify multiple harmonics through the spectrogram. 
For simplicity in our analysis, we choose to select the dominant mode and isolate that contribution to the signal. 
Isolating this track will simplify the corresponding time-domain signal, making it easier to fit the resulting bird-call waveform with a gravitational waveform model. 

We begin by identifying the dominant spectral ridge using a greedy forward-backward algorithm. 
To aid in the identification, we seed the algorithm to begin at the point of maximum spectral energy, $t=t_\text{MSE}$, which corresponds to the loudest point of the bird call, or brightest yellow point in the top panel of Fig. \ref{fig:filtered_spectrogram}. 
Using a greedy algorithm ensures the next step, either forwards or backwards in time, will be to the highest energy bin at that next time step. 
Because the greedy algorithm is set to start at an intermediary time of $t=t_\text{MSE}=-0.0297$ s, we require a forward-backward algorithm to traverse the full time series. 
The forward portion of the greedy algorithm traverses $t > t_\text{MSE}$  while the backwards portion traverses $t < t_\text{MSE}$.
To prevent large jumps in frequency, we bound each step in the greedy algorithm to examining the nearest 30 bins, or approximately 130 Hz. 
This continuity condition helps ensure the algorithm remains on one track in the spectrogram. 

Once the greedy algorithm has constructed a path across the spectrogram corresponding to the loudest track, we smooth this path with a median filter. 
A median filter smooths a set of values by replacing each data point with the median of the data points in its immediate neighborhood, making it robust against any large discontinuities that may appear in a data series. 
We choose a kernel size of 10, which means that each point in the greedy algorithm result is replaced by the median of the 5 data points on either side in the time series. 
This choice maintains the results of the greedy algorithm while eliminating any spurious discontinuities that may have arisen from the greedy algorithm. The smoothed version of the dominant mode isolated by the greedy algorithm appears as a dashed white curve in the top panel of Fig. \ref{fig:filtered_spectrogram}.

Having found the primary track, we apply a Gaussian mask around the track to remove all subdominant tracks from the spectrogram. 
After reaching the peak of the chirp, a clear track is not so straightforward to identify, as seen in the final $\sim0.02$ s of the top panel of Fig. \ref{fig:filtered_spectrogram}. In order to avoid removing important physics which may be captured in this portion of the waveform, we implement a time-dependent width for the mask. The mask width varies from 10 bins early in the signal, where the track is well defined, to 200 bins at $t=0$, via a sigmoid function. 
The result of the applying this Gaussian mask appears in the lower panel of Fig. \ref{fig:filtered_spectrogram}.
This is the signal we apply the inverse STFT to in order to return to the time domain. The filtered waveform signal we aim to map to a BBH configuration is shown in the bottom panel of Fig. \ref{fig:waveforms}.

\subsection{Fitting with \texttt{SEOBNRv5PHM}}
\label{sec:SEOBNRv5PHM}

After filtering the spectrogram for the dominant spectral track, we fit this portion of the signal with a waveform model and measure the system parameters. 
We choose to fit with \texttt{SEOBNRv5PHM}, which is a state-of-the-art effective-one-body (EOB) precessing BBH waveform model \cite{Ramos-Buades:2023ehm}. 
For this work, it is relevant that these waveforms are functions of mass ratio, $q$, and the two dimensionless spins of the component BHs, $\vec{\chi}_i$.

First, we must align the two waveforms at their ``merger" $t_c$, which we define as the time of peak amplitude, which is equivalent to $t_{\rm MSE}$ by construction.
Next, we must choose how we are going to compare the bird call waveform and a GW waveform. 
To accelerate the fitting process, as this is a preliminary work, we choose to fit the amplitude envelope of our signal (red outline of Fig. \ref{fig:waveforms}) with a signal envelope computed with \texttt{SEOBNRv5PHM}. 
Comparing the envelopes motivates the use of a L2 norm residual $\mathcal{R}$ between the two signals, integrated over the entire pre-merger waveform: 
\begin{equation}
\label{eq:residual}
    \mathcal{R}(q,\vec\chi_1,\vec\chi_2)\equiv \int_{-\infty}^{t_c} |h_{\rm EOB}(t;q, \vec{\chi_1}, \vec{\chi_2}) 
    - h_{\rm bird}(t)|^2 dt\,.
\end{equation}

Our optimization function minimizes the residual stated in Eq.~\eqref{eq:residual}.
We utilize the Nelder-Mead method in the \texttt{scipy.minimize()} function to fit a waveform to our signal. 
As suggested by Eq.~\eqref{eq:residual}, we minimize over the mass ratio and the two spin vectors.

\section{Results}
\label{sec: results}

Because this is an exploratory work in gravi-ornithology, our results span a number of interests in the field of gravitational waves. 
In Sec. \ref{sec:fits}, we present the results of fitting the filtered Northern cardinal chirp with the \texttt{SEOBNRv5PHM} waveform model. 
Section \ref{sec:localization} then provides necessary skymap information to perform electromagnetic follow-up on the observed signal. 
In Sec. \ref{sec:glitches}, we migrate to an analysis of additional birds species, studying the similarity of their calls to common glitch morphologies.

\subsection{Waveform fit}
\label{sec:fits}
\begin{figure}
    \includegraphics[width=\linewidth]{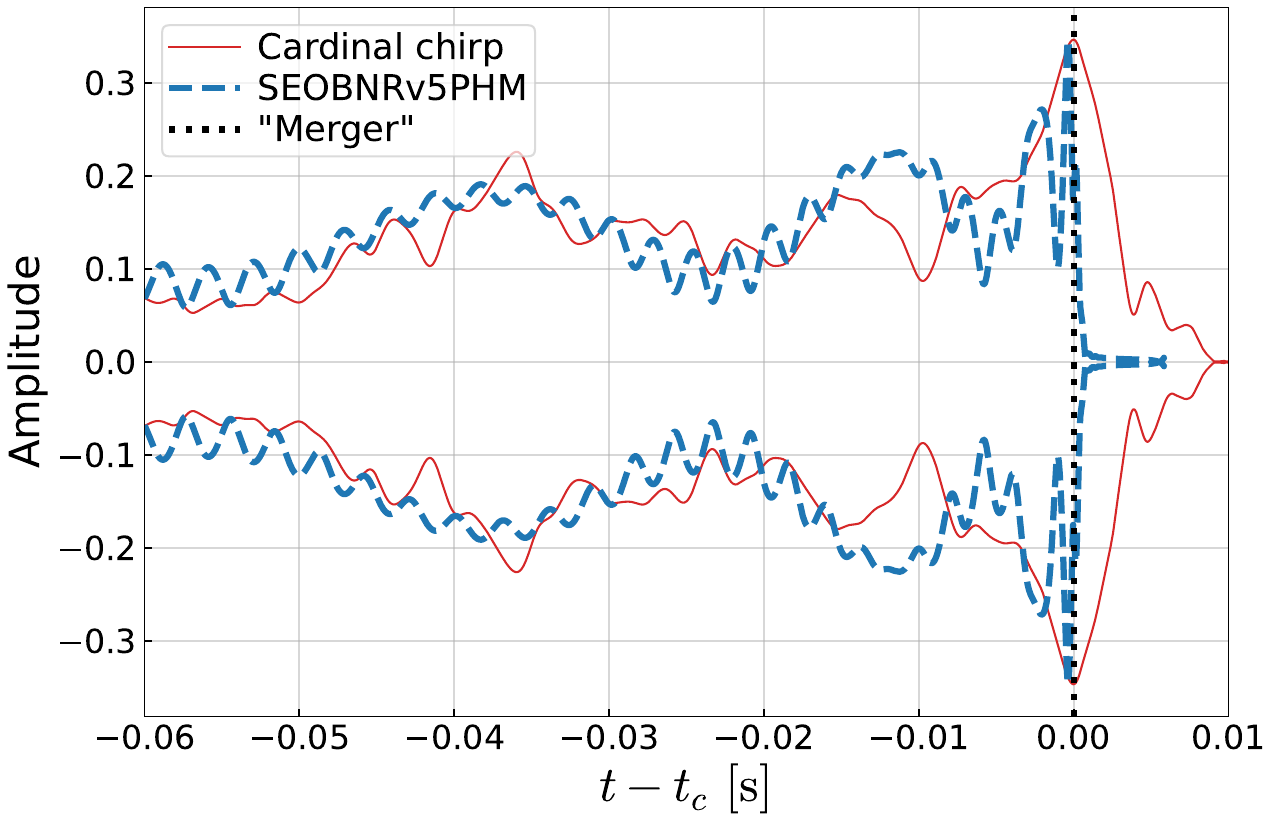}
    \caption{Best fit \texttt{SEOBNRv5PHM} (blue dashed) waveform for the dominant spectral ridge of the Northern cardinal chirp (red solid).
    We restrict the fitting to the modulations beginning at $t\sim -0.25$ s due to noise at early times in the bird chirp inspiral. 
    The best fit corresponds to $q\approx104$, $\chi_{\rm eff} \approx -0.08$, and $\chi_p \approx 0.35$.
    Our exploratory fitting procedure captures some early-time modulations, but the poor fit near the peak amplitude suggests unmodeled physical processes at work.}
    \label{fig:waveform_fit}
\end{figure}

Using the minimization procedure discussed in Sec. \ref{sec:SEOBNRv5PHM}, we fit the dominant harmonic of our signal with a precessing BBH waveform. 
The resulting fit is shown in Fig. \ref{fig:waveform_fit} (blue) and corresponds to an intermediate-mass ratio inspiral (IMRI) with mass ratio $q\approx104$, $\chi_{\rm eff} \approx -0.08$, and $\chi_p \approx 0.35$. Based on these results, \textit{we propose that} Cardinalis cardinalis \textit{may disguise a high mass ratio, precessing, but anti-merging BBH system.}

While the best-fit waveform captures the majority of features in the bird call signal, it fails to fully track the waveform through merger. 
Our optimization procedure likely insufficiently explores the unequal-mass, precessing parameter space, where degeneracies may prevent full coverage of the five-dimensional space. 
A more robust approach than envelope comparison may also be needed to enable finer parameter space exploration. 
Nevertheless, the quality of the fit during the primary modulation ($t \sim -0.25$ s to $t \sim -0.10$ s) suggests that, even if our optimization is not fully robust, effects from ultra-dense matter or exotic beyond the Standard Model physics may be at play in this system, further limiting our ability to fit this signal with a precessing BH waveform model. 
Despite these minor shortcomings, our results serve as a foundational proof-of-concept for modeling avian vocalizations as waveforms from compact binaries.

To make a rough estimate of the total mass of the system, we take advantage of the high mass ratio of the BBH waveform, allowing us to estimate the frequency at merger by twice the orbital frequency at the innermost stable circular orbit \cite{1972Bardeen} (note $c=G=1$ units here):
\begin{equation}
    2\pi f_\text{peak}=2\times \Omega_\text{ISCO}=\frac{2}{ M_\text{tot}\sqrt{r_\text{ISCO}^{3/2}+\chi_\text{eff}}}.
\end{equation}
For a spin of $\chi_\text{eff}\approx-0.08$, this returns the dimensionless $f_\text{peak}\approx0.0233/M_\text{tot}$, or after reinserting units, $f_\text{peak}=4700/(M_\text{tot}/M_\odot)$ Hz. From the lower half of Fig. \ref{fig:filtered_spectrogram}, we see that the peak of the chirp appears at roughly 6500 Hz, corresponding to a total mass of $M_\text{tot}\approx 0.72M_\odot$. 
We will further explore the numerous implications of our parameter estimation results in Sec. \ref{sec: conclusion}.

\subsection{Sky localization}
\label{sec:localization}
Electromagnetic follow-up on gravitational wave events can provide a great deal of information about the progenitors and remnant. Therefore, it is useful to establish a procedure for localizing the source of these waveforms. While this calculation benefits from the fact that avian chirps are observed from distances of meters, rather than hundreds of megaparsecs, establishing a source location for the \textit{Cardinalis cardinalis} chirp is clouded by the sheer number of recorded instances of this waveform.

\begin{figure}
    \includegraphics[width=\linewidth]{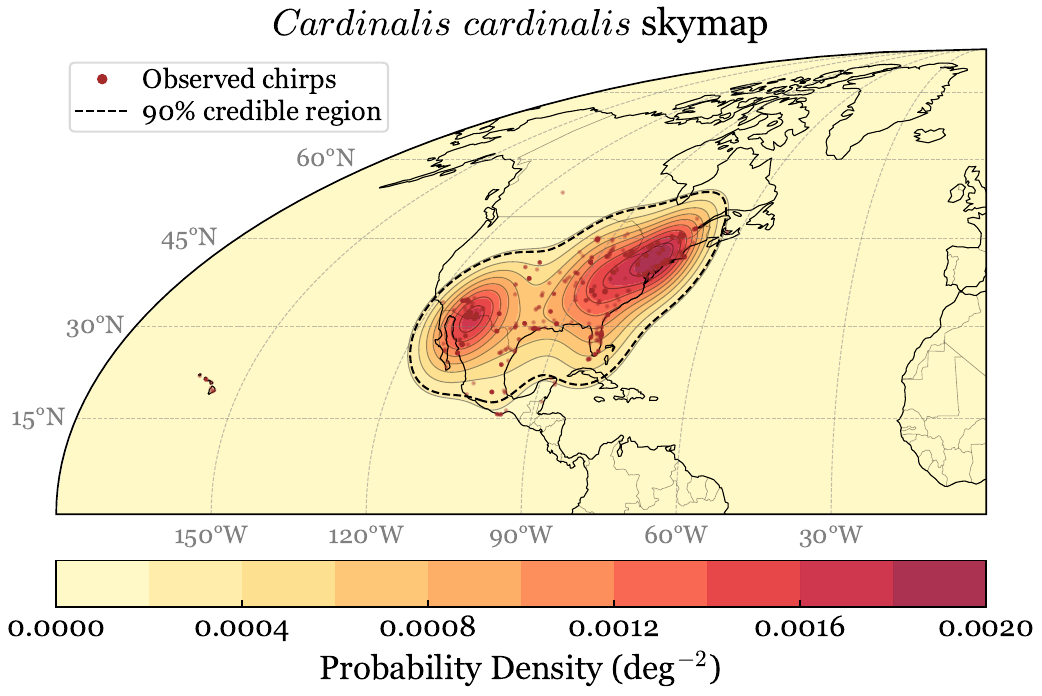}
    \caption{Probability density for the location of the \textit{Cardinalis cardinalis} chirp on the Earth, based on coordinate data from 916 chirp observations (brown dots). The 90\% credible region is denoted by a black dashed contour.
        }
    \label{fig:skymap}
\end{figure}

In Fig. \ref{fig:skymap}, we plot coordinate data from 916 observations of the \textit{Cardinalis cardinalis} chirp and compute a probability density map for the location of this waveform on the Earth. The result is a bimodal distribution featuring two peaks separated by about 30$^\circ$.

The total area captured by the 90\% credible region is $\sim1140\text{ deg}^2$. For comparison, the sky location of GW150914 was reduced to $630\text{ deg}^2$ \cite{LIGOScientific:2016qac}, GW190521 was reduced to between $\sim 800$ and $1200\text{ deg}^2$, depending on the waveform model \cite{LIGOScientific:2020iuh}, and the highly favorable properties of the binary neutron star event GW170817 allowed for reduction to just $16\text{ deg}^2$ \cite{LIGOScientific:2018hze}. Even in this program's infancy, our localization capabilities are roughly on par with those of the advanced LIGO detectors. Electromagnetic follow-up on such signals appears possible, though binoculars may be required. We anticipate that further development of our analysis methods and observations from a growing detector network will enable improved precision in the near future, though localization quality will likely remain quite species-dependent.

\subsection{Glitchy species}
\label{sec:glitches}
Unfortunately, it is not straightforward to map the chirp of any given avian species to a corresponding CBC waveform; in fact, a number of species produce calls which are more reminiscent of glitch morphologies observed in modern-day interferometric GW observatories. 

\begin{figure}
    \includegraphics[width=\linewidth]{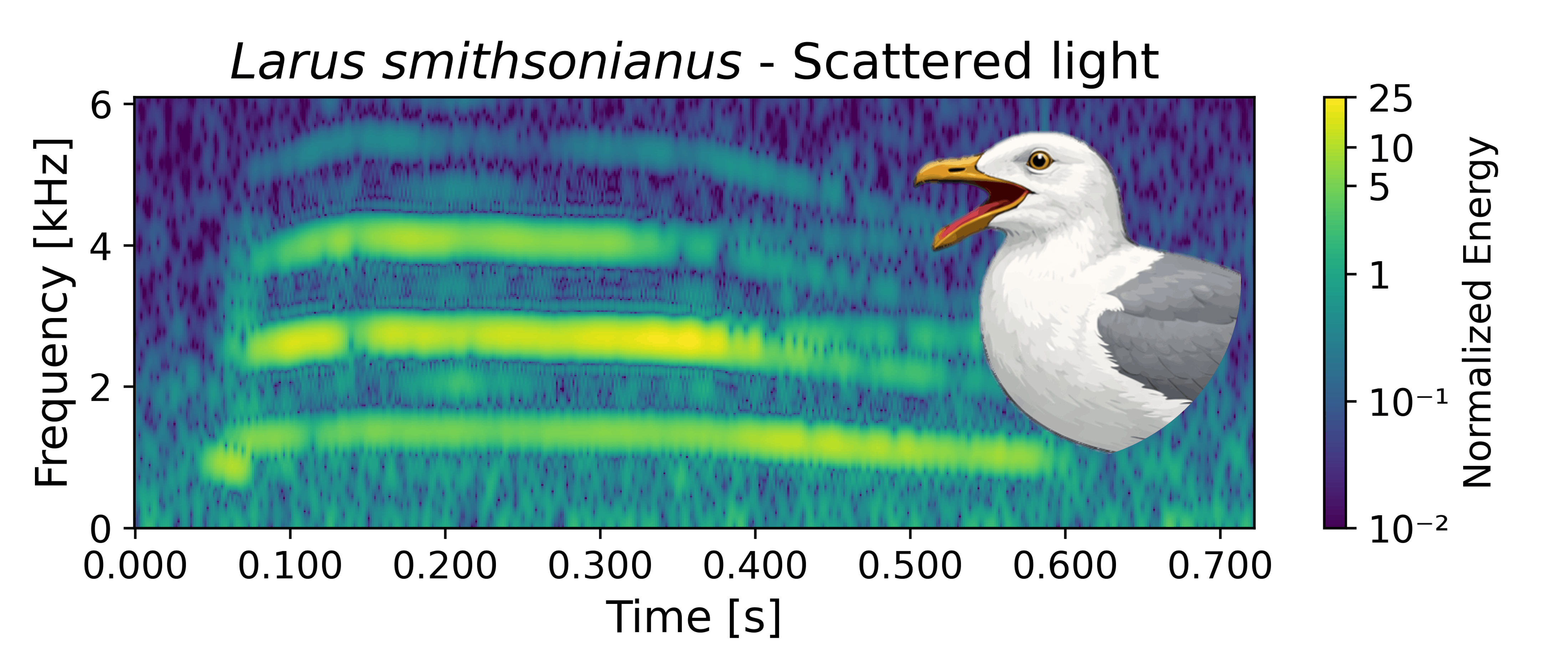}
    \includegraphics[width=\linewidth]{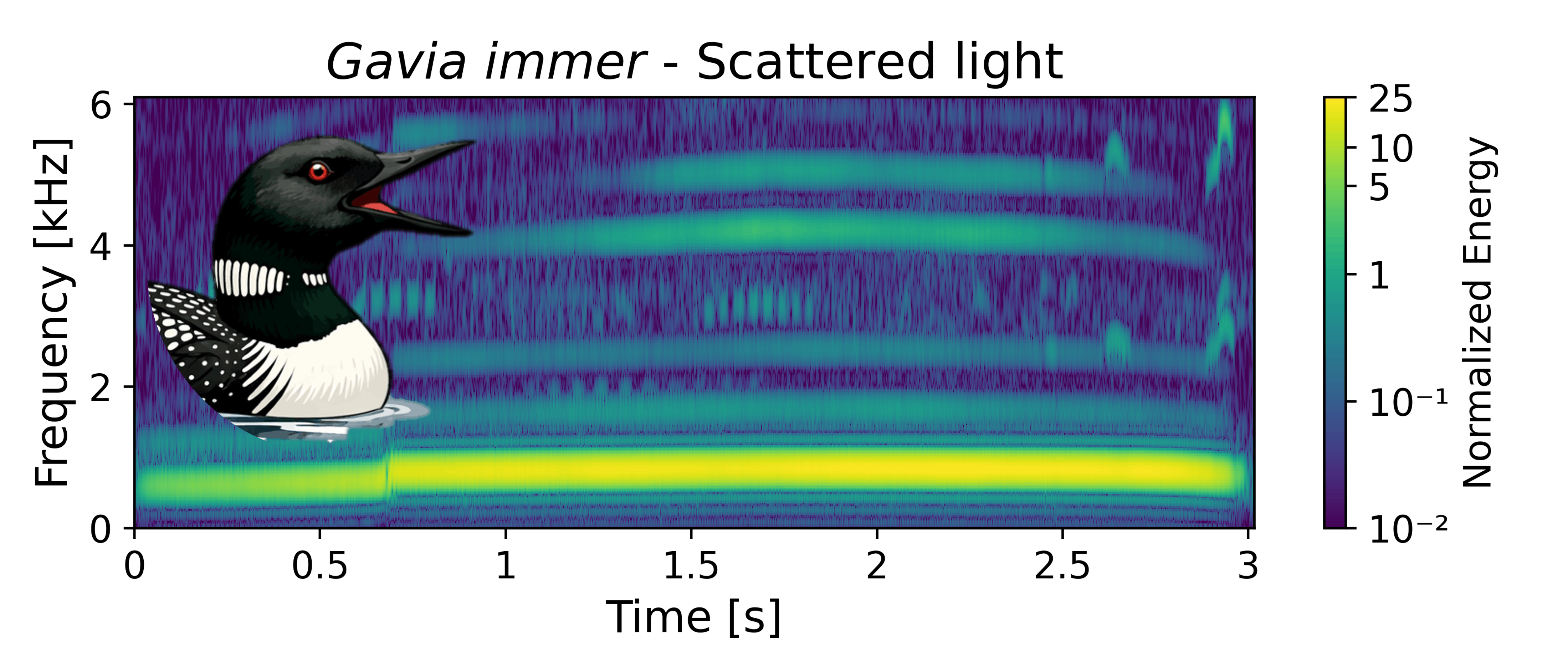}
    \includegraphics[width=\linewidth]{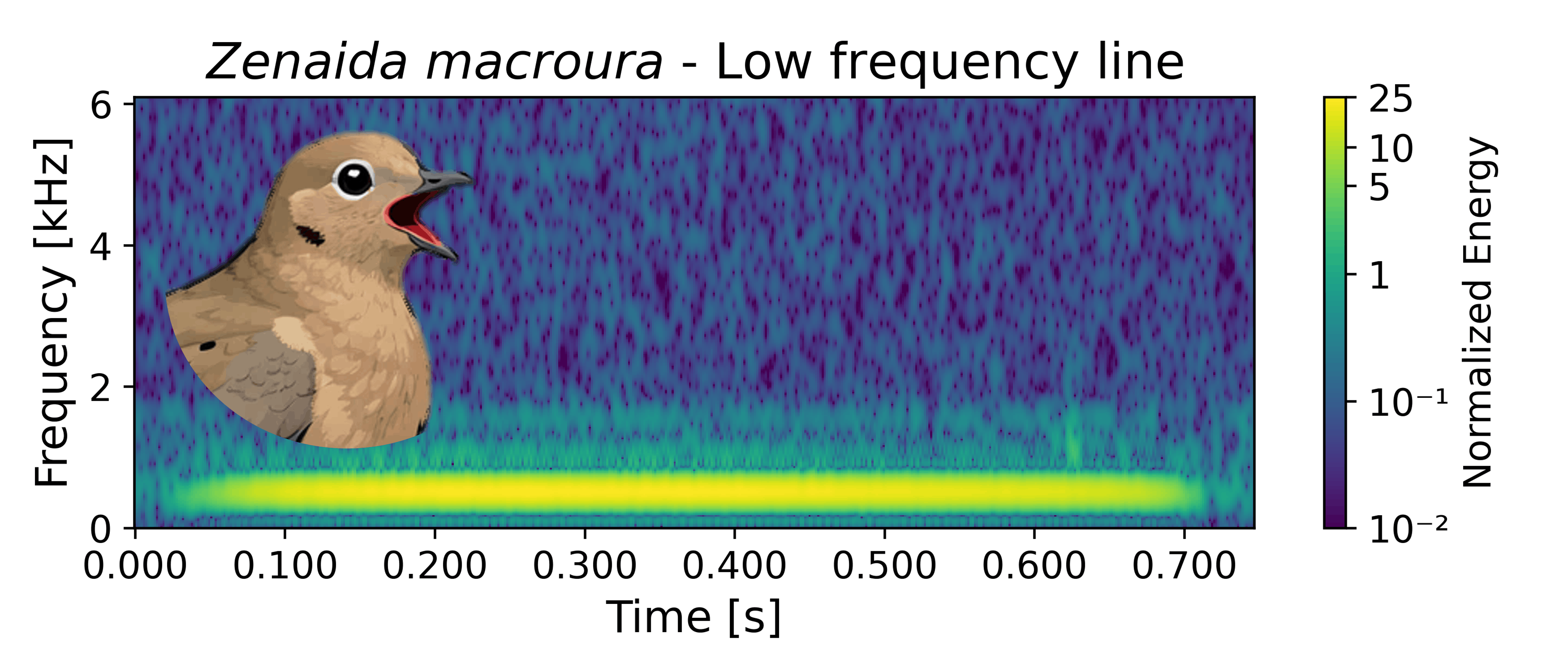}
    \includegraphics[width=\linewidth]{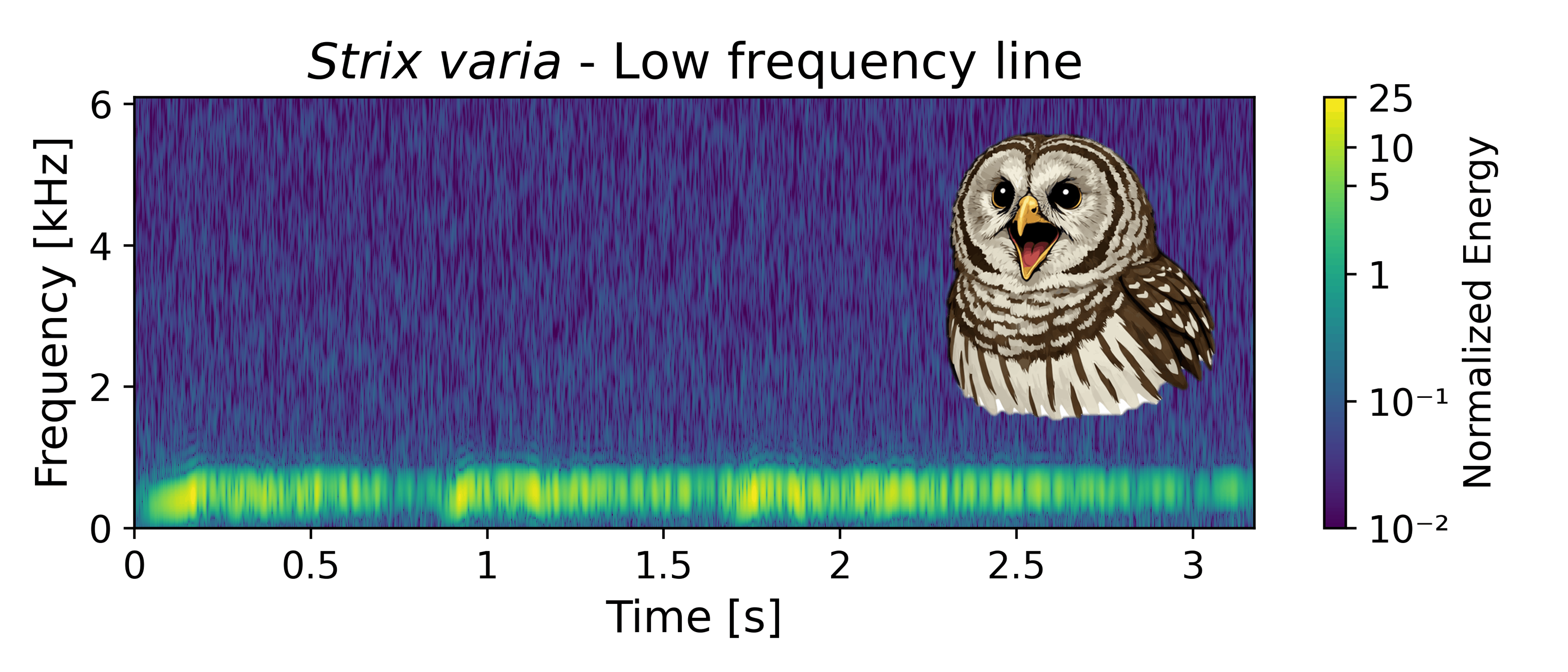}
    \includegraphics[width=\linewidth]{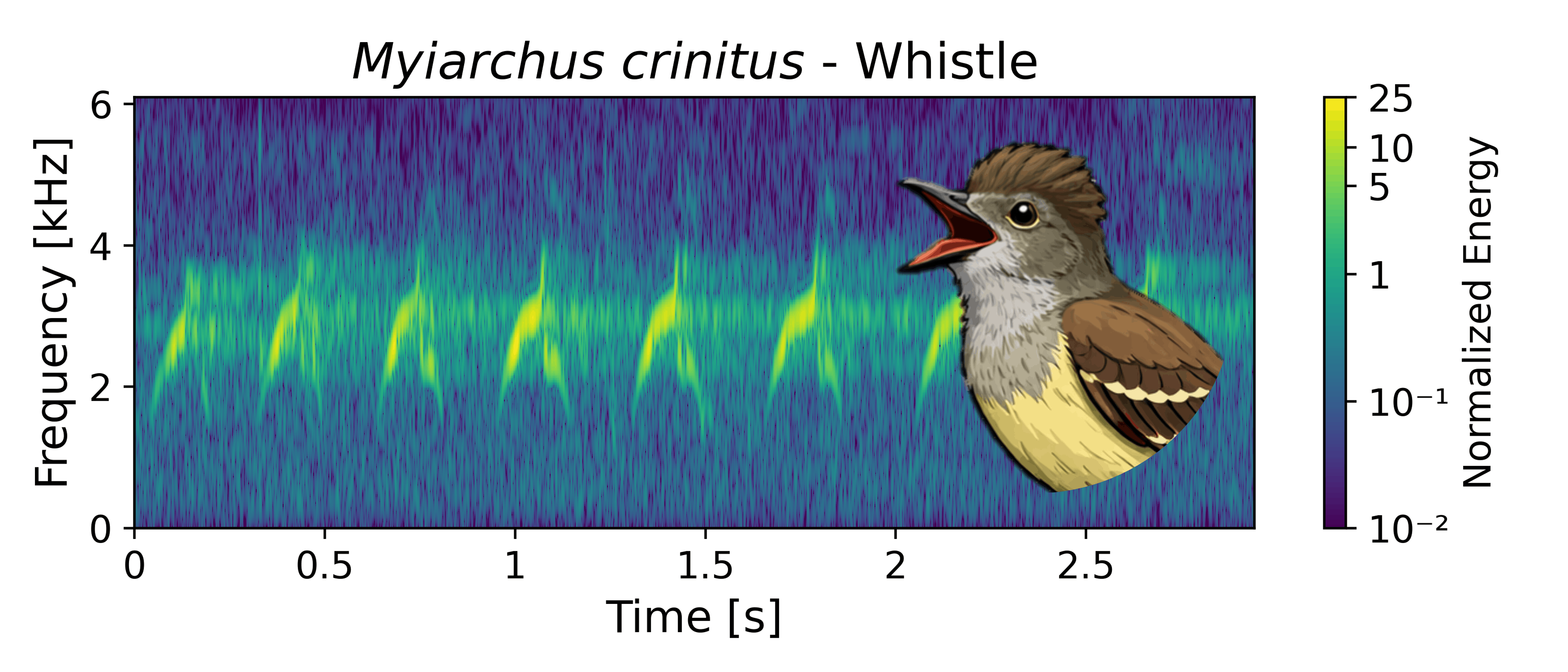}
    \caption{Spectrograms for calls from five choices of bird species, including their electromagnetic counterparts. All are visually comparable to glitches observed in interferometric GW observatories, each of which is noted in the plot titles.}
    \label{fig:glitches}
\end{figure}

As an example, in Fig. \ref{fig:glitches}, we present spectrograms for calls from five different avian species. Each spectrograms bear a passing resemblance to glitches observed in the LIGO detectors and identified by the GravitySpy project \cite{Zevin:2016qwy, Zevin:2023rmt}: light scattering off of optical components in the case of the American herring gull (\textit{Larus smithsonianus}) and the Common loon (\textit{Gavia immer}), low frequency lines from mechanical vibrations in the case of the Mourning dove (\textit{Zenaida macroura}) and the Barred owl (\textit{Strix varia}), and whistles due to beating against radio frequency signals for the Great crested flycatcher (\textit{Myiarchus crinitus}). Determining what causes the chirps of these species to manifest as detector glitches rather than anti-merger waveforms offers an interesting direction for further exploration.

This finding represents an important detail the community should be aware of when conducting further gravi-ornithological studies. Fortunately, both the effects of glitches on GW parameter estimation routines along with procedures for mitigating their impacts have been extensively studied in recent years (see \cite{Davis:2022ird,Payne:2022spz,Udall:2024ovp,Udall:2025bts,Hourihane:2022doe,Hourihane:2025vxc,Ghonge:2023ksb} for just a subset of these works). We anticipate that these ``glitchy species" will not pose an insurmountable roadblock in the continued development of this novel field.

\section{Conclusions \& Discussion}
\label{sec: conclusion}
In this paper, we have completed a basic study on the chirp of the Northern cardinal. 
Using the \texttt{SEOBNRv5PHM} precessing waveform model, we find that we can model the time-reversed chirp as a precessing, unequal mass BBH system. 
Although we are able to extract approximate parameters for the corresponding BBH in the ``inspiral", we fail to match the signal and model at the peak of the waveform. The discovery of the reversed chirp in the \textit{Cardinalis cardinalis} waveform already represents a massive upheaval in our understanding of the behavior of compact objects. Therefore, it is unsurprising that our model fails to generate a perfect match at the peak, as the unmodeled high-energy matter effects or entirely new physics which enable such a system could easily produce more complex anti-merger dynamics than previously anticipated. Despite this slight limitation, we believe that this study encourages continued exploration into gravi-ornithology; extensive statistical analyses of a wide range of avian species are certainly warranted.

Our results have opened a number of urgent new questions: for example, our mass estimate from the \textit{Cardinalis cardinalis} chirp suggests that this species has a primordial origin, as stellar collapse models do not predict the formation of subsolar mass BHs \cite{Hawking:1971ei,Sasaki:2025vql,LIGOScientific:2021job,LVK:2022ydq}. The physical mechanism for generating this species not only must have been active during the early universe, but it also must be capable of making ultra-compact objects of mass $\sim0.7M_\odot$, whose horizon radius should lie near 1 km, fit inside a cardinal of size $\mathcal{O}(\text{cm})$. The discovery of such a mechanism would compose a monumental contribution to our understanding of the behavior of the universe shortly after the Big Bang.

Beyond the origin of these enigmatic creatures, we must also address how birds produce the time-reversed chirps which we have observed here. Are birds able to harness some sort of exotic physics which allows for compact objects to anti-merge, or do they inherently source violations of general relativity itself? Indeed, certain modified theories of gravity have predicted reverse chirp waveforms from core-collapse supernovae \cite{Sperhake:2017itk,Geng:2020slq}, lending credence to the prospect of potential GR violations. Identifying the origin of this behavior may allow for highly accurate species classification purely through waveform parameter estimation or provide insight into the mechanism by which certain species produce glitchy calls rather than reverse chirps.

Finally, birds are known to make multiple chirps, often in rapid succession. Determining the underlying physics behind this phenomenon poses another difficult, yet essential problem to be solved. Do birds actively accrete stellar material throughout their lives which allow for many repeated anti-mergers, or were all the necessary ingredients contained within the egg as it formed in the primordial universe? 
Understanding this process could enable unprecedented tests of solutions to the black hole information paradox \cite{Giddings:1995gd} via careful observations of avian Squawking radiation.

All of these questions are no doubt both clear and pressing; given their relevance to outstanding problems in fundamental physics, we hope they will shape the future directions of this fledgling new field.

\acknowledgements

We thank Artemis and Athena Schwister and Eloise ``Big Mac" Knapp (\textit{Felis catus}), whose feline devotion to the observation of birds inspired this undertaking.
We also acknowledge the wild Red-crowned parrots (\textit{Amazona viridigenalis}) of Pasadena, California for their ritualistic reminder of the volume and frequency of bird calls as well as the fervent Xeno-Canto bird-watching community for their tireless contribution of bird call recording data. Finally, we thank Alan Weinstein for constructive comments provided during the writing of this manuscript.

\bibliography{main}

\end{document}